# Enhancing Expressway Ramp Merge Safety and Efficiency via Spatiotemporal Cooperative Control

Ting PENG*, (Member, IEEE), Xiaoxue XU, Yuan LI, Jie WU, Tao LI, Xiang DONG, Yincai CAI, Peng WU, Sana Ullah

Key Laboratory for Special Area Highway Engineering of Ministry of Education, Chang'an University,710064, Xi'an, China

Corresponding author: Ting Peng (e-mail: t.peng@ieee.org).

This work was supported by the National Natural Science Foundation of China under Grant No.52378430.

**ABSTRACT** In the context of autonomous driving on expressways, the issue of ensuring safe and efficient ramp merging remains a significant challenge. Existing systems often struggle to accurately assess the status and intentions of other vehicles, leading to a persistent occurrence of accidents despite efforts to maintain safe distances. This study proposes a novel spatiotemporal cooperative control approach integrating vehicle-road coordination to address this critical issue. A comprehensive methodology is developed, beginning with the calculation of safe distances under varying spatiotemporal conditions. This involves considering multiple factors, including vehicle speed differentials, positioning errors, and clock synchronization errors. Subsequently, an advanced vehicle conflict risk evaluation model is constructed. By incorporating collision acceleration and emergency acceleration as key parameters, this model offers a more accurate and detailed evaluation of potential risks during the ramp merging process. Based on the calculated safe distances and conflict risk evaluations, a mainline priority coordinated control method is formulated. This method enables the pre-planning of vehicle trajectories, effectively reducing conflicts among vehicles. Through rigorous simulations using diverse traffic volume and speed scenarios, the efficacy of the proposed strategy is validated. The results demonstrate remarkable improvements, with the average delay time reduced by an impressive 97.96% and fuel consumption decreased by 6.01%. These outcomes indicate that the proposed approach not only enhances the speed of vehicle merging but also significantly reduces latency and fuel consumption, thereby enhancing the overall performance of ramp merging operations.

## I. INTRODUCTION

Research on cooperative control strategies in ramp merging areas is a hot topic in the fields of intelligent transportation systems and autonomous driving technologies. With the development of vehicle-to-everything (V2X) communication technology and the increase in intelligent vehicles, many scholars have investigated cooperative merging strategies for intelligent connected and autonomous vehicles to enhance traffic efficiency and safety by optimizing vehicle trajectories and merging sequences. In terms of optimization control strategies, most literature adopt optimal control strategies such as mixed-integers non-linear programming[1],[2] Pseudo-spectral methods[3], game theory[4],[5],[6], hierarchical control strategies[7],[8],[9] distributed control[10],[11],[12] and centralized control [13],[14] to precisely plan the merging behaviors and trajectories of vehicles.

In terms of cooperative control, the literature commonly utilizes cooperative adaptive cruise control[15] and multi-agent systems to coordinate the merging behaviors of multiple vehicles, thereby improving the overall traffic flow. Through V2X communication technology, real-time information sharing and collaborative decision-making are achieved between vehicles and infrastructure and vehicles. In terms of simulation validation, all methods are evaluated for their effectiveness using different simulation platforms such as SUMO and Simulink[16], to assess their performance advantages under various traffic volumes and speed conditions. Through the aforementioned optimization control strategies, cooperative control methods, and simulation validation approaches, these studies demonstrated effective means to enhance vehicle merging efficiency and safety in complex traffic environments.

On the other hand, some studies have employed centralized approaches based on game theory[17] to optimize overall fuel consumption and total travel time by formulating fair and comprehensive game rules to enhance traffic efficiency. Additionally, advanced algorithms such as reinforcement learning[18],[19] and Model Predictive Control have been widely applied for real-time dynamic optimization.

The existing literature has proposed various innovative solutions for studying cooperative control methods in expressway ramp merging areas. Some methods employ optimization scheduling techniques, such as dynamic conflict

graphs[20],[21],[22], which abstract the merging problem as a graph search problem and achieve optimal solutions through heuristic search strategies to reduce overall travel delays. Additionally, some studies have utilized hierarchical system designs[23],[24] including tactical planning and motion planning models, and improved computational efficiency using algorithms such as Monte Carlo tree search to achieve flexible merging positions and a safe and efficient merging process. Queue-based networked autonomous vehicle cooperative optimal control algorithms[25],[26],[27] have been widely discussed. Through the distributed cooperative control of multiple local queues, these algorithms transform complex merging problems into one-dimensional queue-following control problems, thereby improving traffic efficiency. These studies comprehensively utilized graph search, game theory, hierarchical system design, and autonomous driving technology, providing diverse and efficient solutions for cooperative control in expressway ramp merging areas. The risk of collision between two vehicles should not be underestimated. Despite the reduction in traffic accidents owing to autonomous driving, accidents still occur. Therefore, many scholars have assessed the risk of accidents involving autonomous vehicles.

Rahman et al.[28] evaluated the safety of vehicle operation under mixed road conditions with traditional and autonomous vehicles. They used five alternative safety indicators, including speed standard deviation, exposure time collision time, time integral collision time, exposure time rear-end risk index, and side-swipe collision risk. A. Shetty et al.[29] proposed a risk assessment framework that utilizes human driving and road test data to provide insights into the safety of autonomous vehicles. Hu W et al.[30] proposed a collision risk assessment framework based on the prediction of trajectories of other vehicles. This framework integrates solutions such as expected path planning of other vehicles, description of uncertainties in the driving process, and trajectory changes caused by obstacle intrusion. Han et al.[31] proposed a novel spatial-temporal risk field from the perspective of spatiotemporal coupling. This risk field represents the dynamic driving risk of autonomous vehicles in dynamic traffic. M. Jiang et al.[32] proposed a vehicular end-edge-cloud computing framework to facilitate end-edge-cloud vertical cooperation and horizontal cooperation among vehicles. Under the framework, a two-stage reinforcement learning is implemented to obtain the optimal policy for vehicle control. H. Xu et al.[33] designed the terminal machine learning task model and the edge machine learning task model on the vehicle side and road side unit to optimize execution latency, processing accuracy of the cooperative learning system.

In summary, current manual and automated driving technologies still struggle to accurately and promptly acquire the status and driving intentions of the surrounding vehicles. Despite efforts to maintain appropriate safety distances between vehicles on expressway merge zones, a significant number of traffic accidents still occur. Owing to the necessity of maintaining safe distances between vehicles, the capacity for road traffic cannot be further increased, posing a formidable barrier for overcoming traffic congestion. Most scholars calculate the safety risk of autonomous vehicles based on risk indicators or propose new frameworks or models for real-time risk assessment or risk control. However, in these studies, there has been little assessment from scholars regarding the harm and urgency of vehicle collisions.

In this paper, a calculation method for the safe distance between vehicles in the vehicle-road deep cooperation scenario is proposed. Additionally, a quantitative evaluation model for the urgency of vehicle collisions is proposed, based on the distance and speed difference between vehicles. The critical degree of vehicle collision is quantitatively assessed based on the acceleration of the vehicle when the collision occurs. Finally, a collaborative control method is proposed to pre-planned vehicle trajectories to improve vehicle safety and traffic efficiency in on-ramp merge areas of expressways. These findings highlight the significant effectiveness of the proposed spatiotemporal control methods for expressway ramp merging.

The technical roadmap of this paper is shown in Figure 1.

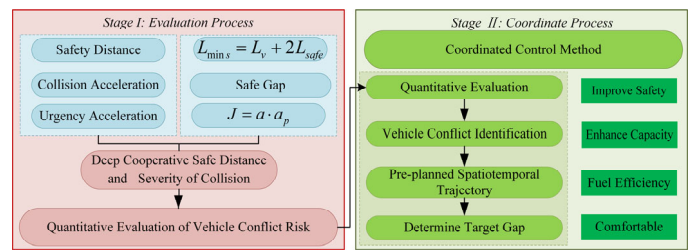

**FIGURE 1.** Technical Roadmap

Our contributions are summarized as follows:

• A calculation method for the safe distance and quantitative evaluation model for the urgency of vehicle collisions are proposed.

• Addressing the expressway ramp merging issue. To proactively resolve the mutual conflicts among vehicles within the ramp merging zone, we propose pre-planning vehicle trajectories to avert spatiotemporal conflicts.

• Our model can reduce average delay time by 97.96% and decrease fuel consumption by 6.01%.

• Our approach effectively addresses issues such as traffic congestion and collision risks in merging zones, thereby making substantial contributions to the development of intelligent transportation systems.

## II. Conflicts Risk Evaluation
### A. Selection of Quantitative Indicators for Conflicts Risk

Given the insufficient research in academia on assessing the severity of collision hazards, this study selected safety distance and conflict urgency as indicators of conflict risk. This selection was based on a series of theoretical foundations and analytical processes, as outlined below:

#### 1) SAFETY DISTANCE

The safety distance refers to the minimum distance[34] that should be maintained to prevent collisions between vehicles. When the distance between vehicles is less than this safety distance, the risk of collision increases significantly. Therefore, the safety distance is an important indicator for assessing conflict risk. Based on the preliminary research, follow-up studies were conducted.

#### 2) CONFLICT URGENCY

Collision acceleration reflects the danger of vehicle collisions. When the acceleration is low, it only affects the passengers' comfort. However, when acceleration is particularly high, it can directly pose lethal problems to passengers. Therefore, collision acceleration is an important indicator for assessing the risk of conflict.

Urgent acceleration reflects the urgency of the collision occurrence. When the urgent acceleration is high, it indicates that the two vehicles are about to collide.

Neither collision acceleration nor urgent acceleration alone can comprehensively reflect the risk of vehicle collisions. Thus, by multiplying them, the degree of conflict urgency is obtained. This combined metric can simultaneously reflect the danger and urgency of vehicle collisions. Therefore, conflict urgency is considered an important indicator of conflict risk.

### B. Deep Cooperative Safety Distance

The safety distance includes three components: the safe distance to be maintained between merging vehicles and mainline vehicles when vehicles enter the mainline, positioning error of the Global Positioning System (GPS), and timing synchronization error between autonomous driving vehicles and the national time synchronization center.

Therefore, the formula for calculating the safe distance to be maintained between the preceding and following vehicles is

$$L_{safe} = 2L_1 + L_2 + L_3, \quad (1)$$

where $L_1$ is the GPS positioning error, measured in meters. $L_2$ represents the distance that is induced by the clock synchronization error between the vehicle's onboard clock and the national time synchronization center, with the measurement unit being meters. $L_3$ is the distance needed to maintain the speed difference between the autonomous driving vehicle and the preceding or following vehicle, measured in meters.

The sources of GPS positioning errors are diverse, such as satellite orbit errors, signal propagation delays, multipath effects, and receiver clock errors. Among these, satellite clock errors and receiver clock errors are the main sources of time synchronization errors and also significant contributors to GPS positioning errors. In this study, the typical value of GPS Real-time Kinematic positioning error is taken as 0.02 meters. Therefore, the positioning error value for expressway sections in this study is: $L_1 = 2cm = 0.02m$.

Hence, it is necessary to calculate the error distance generated during the vehicle's operation based on the clock error between vehicles and the vehicle's operating speed. The formula for this calculation is

$$L_2 = 3 \times 10^{-9} s \cdot (v_1 + v_2), \quad (2)$$

where $v_1$ is the driving speed of the preceding autonomous driving vehicle, measured in meters per second (m/s). $v_2$ is the driving speed of the following autonomous driving vehicle, measured in meters per second (m/s).

When the speed of the following vehicle is greater than that of the leading vehicle, the formula for calculating the safe distance retained for the speed difference between the two vehicles is

$$L_3 = \frac{(v_1 - v_2)^2}{254(\phi + \varphi)}, \quad (3)$$

where $L_3$ is the safe distance retained for the speed difference between the two vehicles, measured in meters. $v_1$ is the traveling speed of the following vehicle, measured in kilometers per hour (km/h). $v_2$ is the traveling speed of the leading vehicle, measured in kilometers per hour (km/h). $\phi$ is the coefficient of adhesion between the road surface and the tires, taken as 0.4. $\varphi$ is the road resistance coefficient, taken as 0.11.

When merging in the merging area, it's necessary to maintain a sufficient safety gap between the two mainline vehicles to allow the merging vehicle from the entrance ramp to merge safely and smoothly onto the mainline. The minimum required safety gap is twice the safe distance that should be maintained between the preceding and following vehicles, plus the sum of the lengths of the merging vehicle's body. When calculating the safe distance retained for the speed difference and clock precision error, the speeds of the two vehicles were taken as those of the merging and mainline vehicles, respectively. The formula for this calculation is as follows:

$$L_{\min s} = L_v + 2L_{safe}, \quad (4)$$

where $L_{\min s}$ is the minimum safe gap left between the mainline vehicles when the merging vehicle from the entrance ramp merges into the mainline, measured in meters. $L_v$ is length of the body of the vehicle merging onto the mainline, measured in meters. $L_{safe}$ is the safe gap between the merging vehicle and the mainline vehicle at the moment of merging, measured in meters.

### C. The Severity of Conflict

A single indicator, neither collision acceleration nor urgent acceleration, can fully represent the severity of conflict. Hence, multiplying the collision acceleration by urgent acceleration yields the collision severity. The formula for the calculation is as follows:

$$J = a \cdot a_p, \quad (5)$$

where $J$ is severity of potential collision between vehicles. $a$ is collision acceleration, measured in meters per second squared (m/s²). $a_p$ is urgency acceleration, measured in meters per second squared (m/s²).

When the collision acceleration is high and the urgency level is high, the resulting collision severity for the vehicle is high. However, if the collision acceleration is high but the urgency level is low, or if the collision acceleration is low but the urgency level is high, the resulting severity of the collision for that vehicle may not be high.

#### 1) COLLISION ACCELERATION

Let the speed difference between the two vehicles be denoted as $\Delta v$, ranging from 0 to 100 m/s, which corresponds to 0 to 360 km/h, covering almost all possible collision speed ranges. The maximum speed difference occurs when two vehicles collide head-on, where, because the velocity is a vector, the speed difference is the sum of the scalar velocities of the two vehicles. Let $k = \frac{m_1}{m_2}$ represent a certain value, and let the masses of the two colliding vehicles be denoted as $m_1$ and $m_2$. The relationship between the masses $m_1$ and $m_2$ of the vehicles is denoted as $m_1 = km_2$. In this study, the range of $k$ was from 0.001 to 1000, where values of 0.001 and 1000 represent extreme cases. Diagrams illustrating the vehicles before and after the collision are shown in Figure 2 and Figure 3 respectively.

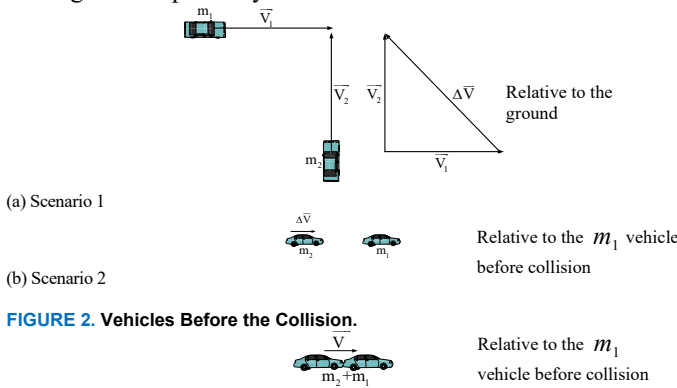

(a) Scenario 1

(b) Scenario 2

**FIGURE 2.** Vehicles Before the Collision.

**FIGURE 3.** Vehicles After the Collision.

Owing to the wide range of $k$, spanning several orders of magnitude, directly plotting it would result in insignificant distinctions between small values. However, using a logarithmic coordinate axis can address this issue. Additionally, employing a logarithmic coordinate axis facilitates easier observation of patterns and trends across the entire data range. Let $p = \lg k$, then $k = 10^p$, so $p_{\min} = \lg 10^{-3} = -3$, and $p_{\max} = \lg 10^3 = 3$. Consequently, the range of $p$ values is from -3 to 3.

In vehicles before the collision, scenario 1 uses the geodetic reference system, while scenario 2 uses the coordinate system of the $m_1$ vehicle as the reference frame. The vehicles after the collision are shown in Figure 3, in which the $m_1$ vehicle before the collision is used as the reference frame. According to the relationship for a completely inelastic collision, the velocity ($\overline{V}$) can be determined as $\overline{V} = \frac{m_2}{m_1 + m_2} \cdot \overline{\Delta v}$.

In the scenario where a vehicle with mass $m_1$ traveling uniformly in a straight line before the collision, with its coordinate system serving as the reference frame, this implies that the vehicle with mass $m_1$ is stationary. Meanwhile, another vehicle with mass $m_2$ travels at velocity $v_2$ towards it, resulting in a collision.

When two vehicles collide, the momentum is conserved both before and after the collision, thus satisfying the following equation:
$$m_2 v_2 = (m_1 + m_2) \cdot a \cdot t, \quad (6)$$

where $v_2$ is the velocity of the vehicle with mass before the collision, measured in meters per second (m/s). $a$ is the acceleration generated during the collision of two vehicles, measured in meters per second squared (m/s²). $t$ is duration from the beginning to the end of the collision between two vehicles, measured in meters.

The danger to vehicles after a collision is related to the collision acceleration that occurs afterward. The smaller the acceleration of the vehicle after the collision, the safer it is, whereas a higher acceleration poses greater danger.

Because the reference frame is established based on a vehicle with mass $m_1$, which is assumed to be moving uniformly along a straight line, it appears stationary in this frame of reference, with its velocity assumed to be $v_1 = 0$. Hence, the speed difference between the two vehicles was $|\overrightarrow{\Delta v}| = |\overrightarrow{v_2} - \overrightarrow{v_1}| = |\overrightarrow{v_2}|$ (m/s)

Substituting $|\overrightarrow{\Delta v}| = |\overrightarrow{v_2}|$ and $m_1 = km_2$ into equation (3), the formula for calculating the average acceleration during the collision is obtained as

$$a = \frac{|\overrightarrow{\Delta v}|}{(1+k) \cdot t}, \quad (7)$$

where the collision time t is taken as 0.2 seconds, substituting $k = 10^p$ into equation (4), the formula for calculating the collision acceleration is obtained as

$$a = \frac{5|\overrightarrow{\Delta v}|}{1 + 10^p}, \quad (8)$$

where $a$ is collision acceleration, measured in meters per second squared (m/s²). $|\overrightarrow{\Delta v}|$ is speed difference between the two vehicles, measured in meters per second (m/s).

These cover all scenarios of collisions between vehicles, regardless of whether the vehicles have larger or smaller masses or higher or lower speeds. This is because the chosen reference frame is that of the vehicle moving at constant velocity before the collision, where the vehicle is stationary, and its velocity is zero. If a collision occurs between two vehicles, it is because the other vehicle is in motion. The velocity of the other vehicle relative to the stationary vehicle, which is traveling at constant velocity, ranges from 0 to 100 m/s, covering all possible speeds. Similarly, the ratio of the mass of the stationary vehicle to that of the other vehicle ranges from 0.001 to 1000, covering all possible scenarios of vehicle masses.

After the collision, the vehicle that was initially stationary in uniform rectilinear motion experiences a change in velocity relative to the original reference frame. Its velocity changes from zero to $\frac{|\overrightarrow{\Delta v}|}{1+k}$. To simplify the matter, this process is modeled as an inelastic collision. Under the assumption that the two vehicles do not separate after the encounter, they will both move forward at the same velocity $\frac{|\overrightarrow{\Delta v}|}{1+k}$. In different inertial reference frames, the velocities of the vehicles differ, but the accelerations are the same. For collisions, the greater the acceleration, the more dangerous it is. Therefore, only the study of acceleration is needed, not the magnitude of the velocity.

The acceleration generated during collision depends not only on the mass of the vehicles but also on the difference in their velocities. The abscissa and ordinate represent the velocity difference and the logarithm of the mass ratio, respectively, using a logarithmic scale of base 10. Because the range of the mass ratio spans multiple orders of magnitude, directly plotting it would obscure the differences between small values. Using a logarithmic scale allowed us to observe trends across the entire data range more easily. When $p = 0$, indicates equal masses for both vehicles, $m_1 = m_2$, the acceleration during collision is primarily determined by the velocity difference. As shown in the Figure 4, a greater velocity difference leads to a higher acceleration during collision, resulting in increased danger to the vehicles.

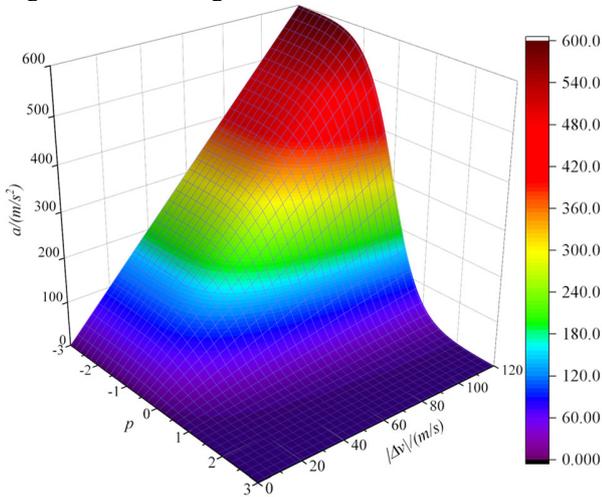

**FIGURE 4.** Relationship Among $a$, $|\Delta v|$ and $p$.

### 2) URGENT ACCELERATION
However, relying solely on acceleration at the moment of collision may not fully capture the collision peril. For instance, if a collision occurs in the next second, there would be insufficient time to adjust the vehicle's trajectory, resulting in a high level of urgency. Conversely, if the collision occurred 30s later, there would be less urgency to adjust the vehicle's trajectory. Therefore, urgent acceleration is chosen to represent the urgency of the collision, considering the timing of the collision occurrence.

Given that the distance between two cars is $S$ and the velocity difference between them is $\Delta v$, the acceleration of a vehicle traveling from a higher speed to match the velocity of the preceding vehicle is

$$a_p = \frac{v_1^2 - v_2^2}{2S}, \qquad (9)$$

where $a_p$ is urgent acceleration, measured in meters per second squared (m/s²). $v_1$ is speed of faster vehicle, measured in meters per second (m/s). $v_2$ is speed of slower vehicle, measured in meters per second (m/s). $S$ is the distance between the two vehicles, measured in meters.

This acceleration represents the urgency of a collision between vehicles and is called the urgent acceleration. A smaller urgent acceleration indicates that the required adjustment in speed for the vehicles to match velocities is less within a given unit of time, resulting in a lower level of urgency for the collision. Conversely, a larger urgent acceleration implies that a greater adjustment in speed is needed within the same unit of time, indicating a higher level of urgency for the collision.

## III. Merging Area Control
### A. The judgment process for coordinated control
Non-coordinated control refers to the free movement of both mainline and ramp vehicles. In this study, the free movement state of vehicles is defined as follows: mainline vehicles travel at a constant speed of $v_0$, whereas ramp vehicles travel at a constant speed of $v_{R0}$ on the ramp until the end of the ramp, then accelerate at a rate of $a_r$ on the acceleration lane to merge directly into the mainline at the same speed as the mainline vehicles.

The initial position of the mainline vehicle is $x_0$, and the functional relationship between station number and time of the mainline vehicle is

$$S_{main} = x_0 + \frac{v_0}{3.6}t. \qquad (10)$$

The movement of ramp vehicles comprises three stages: the first stage is constant-speed driving, the second stage is uniform acceleration, and the third stage is constant-speed driving.

The first stage involves the vehicle traveling at a constant speed $v_{R0}$ from its initial station $r$ until it reaches station 0 (the end of the ramp). Let $t_1 = \frac{|r|}{v_{R0}/3.6}$ denote the time required for a vehicle to travel at a constant speed from its initial station to station 0. During this time, the functional relationship between the station number of the ramp vehicle and time is

$$S_1(t) = r + \frac{v_{R0}}{3.6}t \, (0 \leq t < t_1). \qquad (11)$$

The second stage involves the acceleration of the vehicle with acceleration $a_r$ on the acceleration lane from station 0 until it reaches the same speed as the mainline traffic, then merges directly into the mainline. Let $t_2 = \frac{v_0 - v_{R0}}{3.6 a_r}$ denote the time at which the ramp vehicle accelerates uniformly. During this time, the functional relationship between the station number of the ramp vehicle and time is given by:

$$S_2(t) = S_1(t_1) + \frac{v_{R0}}{3.6}(t - t_1) + \frac{1}{2}a_r(t - t_1)^2 \, (t_1 \leq t \leq t_1 + t_2) \qquad (12)$$

The third stage involves the ramp vehicle traveling at a constant speed $v_0$ after merging into the mainline. During this time, the functional relationship between the station number of the ramp vehicle and time is

$$S_3(t) = S_2(t_1 + t_2) + \frac{v_0}{3.6}(t - t_1 - t_2)(t > t_1 + t_2). \qquad (13)$$

The functional relationship between the station number of the ramp vehicle and time is expressed as equation (14).

$$S_{ramp}(t) = \begin{cases} r + \dfrac{v_{R0}}{3.6}t, & 0 \leq t < t_1 \\ \dfrac{v_{R0}}{3.6}(t-t_1) + \dfrac{1}{2}a_r(t-t_1)^2, & t_1 \leq t \leq t_1 + t_2 \\ S_2(t_1+t_2) + \dfrac{v_0}{3.6}(t-t_1-t_2), & t > t_1 + t_2 \end{cases} \quad (14)$$

In this state of free movement, two main scenarios are likely to occur:
(1) When ramp vehicles can safely merge into the mainline, there is no conflict between the ramp and mainline vehicles, thus no adjustment of the vehicles is required. When vehicles can merge into the mainline safely without coordination, they incur minimum cost, as no vehicle needs to be adjusted.
(2) When conflicts arise between ramp and mainline vehicles, the coordinated control of both ramp and mainline vehicles is necessary. This paper proposes two methods of coordinated control: mainline priority and ramp priority coordinated control methods. Through these two coordinated control methods, ramp vehicles can be safely merged into the mainline. Both these coordinated control methods incur certain costs when ensuring the safe merging of ramp vehicles into the mainline. These costs mainly involve sacrificing the vehicle travel speed or increasing the vehicle fuel consumption.

## IV. Mainline Priority Coordinated Control Method
### 1) SAFE DISTANCE BETWEEN VEHICLES
When the speed of the mainline vehicle is $v_0$ and the speed of the ramp vehicle is $v_{R0}$, even if the ramp vehicle accelerates to match the mainline speed, there is still a speed difference between the mainline and ramp vehicles because the ramp vehicle has both lateral and longitudinal speeds during merging. Assuming that the angle of inclination during merging for the ramp vehicle is $30°$, a schematic of the merging angle for the ramp vehicle is shown in Figure 5. The lateral speed of the ramp vehicle was $v_{Rx} = v_0 \cos 30°$ km/h. Therefore, a speed difference exists between the ramp and the mainline vehicles.

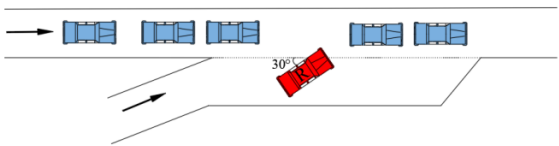

**FIGURE 5.** Vehicle Merging Angle.

Assuming that when the ramp vehicle merges into the mainline, the mainline vehicle in front of the target gap is denoted as X, with its initial station number as $S_X(t=0) = x$; and the mainline vehicle behind the target gap is denoted as Y, with its initial station number as $S_Y(t=0) = y$. During merging, the safe distance between the ramp vehicle and the vehicle in front of the target gap should be

$$L_{r-X} = 2L_1 + \frac{(v_{Rx} - v_X)^2}{254(\phi + \varphi)}. \quad (15)$$

During merging, the safe distance between the ramp vehicle and the vehicle behind the target gap should be

$$L_{r-Y} = 2L_1 + \frac{(v_{Rx} - v_Y)^2}{254(\phi + \varphi)}, \quad (16)$$

where $L_{r-X}$ is safe distance between the ramp vehicle and the front vehicle of the target gap, measured in meters. $L_{r-Y}$ is safe distance between the ramp vehicle and the rear vehicle of the target gap, measured in meters. $L_1$ is GPS positioning error, taken as 0.02 meters, $v_X$ is driving speed of the front vehicle of the target gap when the ramp vehicle merges into the mainline, measured in kilometers per hour (km/h). $v_Y$ is driving speed of the rear vehicle of the target gap when the ramp vehicle merges into the mainline, measured in kilometers per hour (km/h). $v_{Rx}$ is lateral speed of the ramp vehicle when merging into the mainline, measured in kilometers per hour (km/h). $\phi$ is the coefficient of adhesion between the road surface and tires was taken as 0.40. $\varphi$ is the coefficient of road resistance is taken as 0.11. The safe distance between mainline vehicles traveling at the same speed is $L_{m-m} = 0.04m$.

When the ramp vehicle merges into the mainline, the minimum gap that needs to be left between the two vehicles on the mainline is equal to the length of the ramp vehicle plus the safe distance between the ramp vehicle and the vehicles in front and behind. The calculation formula is

$$L_{\min s} = L_v + L_{r-X} + L_{r-Y}, \quad (17)$$

where $L_{\min s}$ is the minimum gap that should be left between the ramp vehicle and the vehicles in front and behind when the ramp vehicle merges into the mainline, measured in meters. $L_v$ is the length of the vehicle body, taken as 5m. $L_{r-X}$ is safe distance that should be maintained between the ramp vehicle and the vehicle in front of the target gap, measured in meters. $L_{r-Y}$ is safe distance that should be maintained between the ramp vehicle and the vehicle behind the target gap, measured in meters.

### 2) CALCULATING THE SEVERITY OF CONFLICT
When the acceleration is less than 0.3g, passengers are in a relatively comfortable state. Therefore, the threshold for the severity of conflict is set to $J = (0.3g)^2 = 0.09g^2$ (where g = 9.8 m/s$^2$).

### 3) VEHICLE CONFLICT IDENTIFICATION
The time for the ramp vehicle to travel at a constant speed on the ramp is $t_1$, $t_1 = \dfrac{|r|}{v_{R0}/3.6}$, whereas the time for it to accelerate uniformly on the acceleration lane is $t_2$, $t_2 = \dfrac{v_0 - v_{R0}}{3.6 a_r}$. When it reaches the same speed as the mainline vehicle in the acceleration lane, denoted as time $t_1 + t_2$, the ramp vehicle merges into the mainline. At this point, the position of the ramp vehicle is

$$S_R(t_1+t_2) = \frac{v_{R0} t_2}{3.6} + \frac{1}{2} a_r t_2^2, \quad (18)$$

where $S_{ramp}(t_1+t_2)$ is milepost position of the ramp vehicle at time $t_1 + t_2$, measured in meters. $a_r$ is the acceleration of the ramp vehicle on the acceleration lane, m/s$^2$. $t_1$ is time taken for the ramp vehicle to travel at a constant speed from the initial milepost to milepost 0, measured in meters. $t_2$ is the time it takes for the ramp vehicle to accelerate from milepost 0 to the point of merging with the mainline vehicle (i.e., where their speeds are the same), measured in meters.

The milepost position of the mainline vehicle at time $t_1 + t_2$ is

$$S_{main}(t_1 + t_2) = x_0 + \frac{v_0}{3.6}(t_1 + t_2). \quad (19)$$

When the milepost positions of the ramp vehicle and the mainline vehicle at time $t_1 + t_2$ satisfy equation (20), a conflict will occur at the merging point.

$$S_{main}(t_1 + t_2) - L_v - 0.04 \leq S_R(t_1 + t_2) \leq S_{main}(t_1 + t_2) + L_v + 0.04 \quad (20)$$

If there are two mainline vehicles that satisfy this equation, the vehicle with the larger milepost position will be selected as the vehicle that conflicts with the ramp vehicle.

### 4) DETERMINE THE TARGET GAP

A gap greater than or equal to the minimum acceptable merging gap is $L_{\min s}$ for the ramp vehicle as the target merging gap. No adjustment is required for the vehicles before and after the selected gap, allowing the ramp vehicle to merge directly.

I Select the gap before or after the mainline vehicle conflicting with the ramp, which is greater than or equal to the minimum acceptable merging gap $L_{\min s}$, is selected as the target merging gap for the ramp vehicle.

The gaps between a vehicle that may collide with the ramp vehicle and its preceding and following vehicles are denoted as $S_{front}$ and $S_{behind}$, respectively. After the sizes of these two gaps were calculated, they were compared.

When $S_{front} \geq L_{\min s}$ and $S_{behind} \geq L_{\min s}$ are valid, if $S_{front} > S_{behind}$ or $S_{front} = S_{behind}$, select the gap between the vehicle that might collide with the ramp vehicle and its preceding vehicle as the target gap. If the ramp vehicle's acceleration is adjusted to merge into this gap; If $S_{front} < S_{behind}$, select the gap between the vehicle that might collide with the ramp vehicle and its following vehicle as the target gap.

II Select a gap greater than or equal to the minimum acceptable merging gap $L_{\min s}$ for the ramp vehicle outside the preceding and following vehicles of the mainline vehicle conflicting with the ramp as the target merging gap for the ramp vehicle.

If the gaps before or after the mainline vehicle conflicts with the ramp are both smaller than the minimum acceptable merging gap $L_{\min s}$, then search for a gap closer to the mainline vehicle that is greater than or equal to the minimum acceptable merging gap $L_{\min s}$ as the target merging gap for the ramp vehicle.

The ramp vehicle's acceleration and merging point must both meet the requirements to merge into the selected target gap greater than or equal to the minimum acceptable merging gap $L_{\min s}$. If the ramp vehicle cannot satisfy both of these requirements simultaneously, then a gap smaller than the minimum acceptable merging gap $L_{\min s}$ is selected the target merging gap for the ramp vehicle.

If the ramp vehicle is accelerating only on the acceleration lane, and given that the length of the acceleration lane is only 200 m, the condition that the acceleration must satisfy is

$$a_r \geq \frac{(\frac{v_0}{3.6})^2 - (\frac{v_{R0}}{3.6})^2}{2S_a}, \quad (21)$$

Where $a_r$ is acceleration of the ramp vehicle on the acceleration lane, measured in meters per second squared (m/s²). $v_0$ is driving speed of the mainline vehicle, measured in kilometers per hour (km/h). $v_{R0}$ is driving speed of the ramp vehicle, measured in kilometers per hour (km/h). $S_a$ is the length of the acceleration lane, taken as 200 meters, measured in meters.

In addition, to ensure passenger comfort and prevent discomfort caused by excessive acceleration, the maximum acceleration is constrained to not exceed 6 m/s². Therefore, the final condition that the acceleration must satisfy is

$$\frac{(\frac{v_0}{3.6})^2 - (\frac{v_{R0}}{3.6})^2}{2S_a} \leq a_r \leq 6 \ m/s^2. \quad (22)$$

The duration of the first phase during which the ramp vehicle travels at a constant speed is $t_1 = \frac{|r|}{v_{R0}/3.6}$. The duration of the second phase, during which the vehicle accelerates uniformly, is $t_2 = \frac{v_0 - v_{R0}}{3.6 a_r}$. After the end of the second phase (uniform acceleration), the requirement for the milepost of the ramp vehicle to satisfy the merging point adjustment is

$$0 < S_R(t_1 + t_2) = \frac{v_{R0} t_2}{3.6} + \frac{1}{2} a_r t_2^2 < 200 \ m. \quad (23)$$

### 5) MAINLINE PRIORITY COORDINATED CONTROL

**(a)** Choosing a gap greater than or equal to the minimum acceptable merging gap $L_{\min s}$ as the target merging gap for the ramp vehicle to merge into.

In this scenario, the ramp vehicle accelerates to merge into the mainline when its speed matches that of the mainline vehicle. The acceleration of the ramp vehicle needs to be determined by formulating equations based on the driving conditions of the mainline vehicle.

The duration of the first phase during which the ramp vehicle travels at a constant speed is

$$t_1 = \frac{|r|}{v_{R0}/3.6}. \quad (24)$$

The duration of the second phase of uniform acceleration is

$$t_2 = \frac{v_0 - v_{R0}}{3.6 a_r}, \quad (25)$$

where $t_1$ is duration of the first phase during which the ramp vehicle travels at a constant speed, measured in meters. $t_2$ is duration of the second phase during which the ramp vehicle accelerates uniformly, s; $r$ is initial milepost of the ramp vehicle, m; $v_0$ is driving speed of the mainline vehicle, measured in kilometers per hour (km/h). $v_{R0}$ is driving speed of the ramp vehicle, measured in kilometers per hour (km/h). $a'_r$ is the adjusted acceleration of the ramp vehicle, measured in meters per second squared (m/s²).

First, select a gap before or after the mainline vehicle conflicting with the ramp that is greater than or equal to the minimum acceptable merging gap $L_{\min s}$ as the target merging gap for the ramp vehicle to merge into.

Next, the spatiotemporal diagram depicts the changes in milepost over time for seven mainline vehicles and one ramp vehicle. The mainline vehicles all travel at a speed of 100 km/h, whereas the ramp vehicle travels at 60 km/h on the ramp and accelerates at 2 m/s² on the acceleration lane until merging directly into the mainline when its speed matches that of the mainline vehicles.

Calculate the gaps between the vehicle that may collide with the ramp vehicle and its preceding and following vehicles, denoted $S_{front}$ and $S_{behind}$, respectively. After calculating the size of these two gaps, we compared them.

When $L_{\min s} \leq S_{behind} \leq S_{front}$ is established, the target gap between the vehicle intending to merge and the vehicle in front of it, which may collide with the merging vehicle must be selected. The acceleration of the movement of the merging vehicle should be adjusted to merge with this gap.

When $L_{\min s} \leq S_{front} < S_{behind}$ is established, the gap between the vehicle that is about to collide with the merging vehicle and the vehicle behind it was selected as the target gap;

After selecting the target gap, the vehicle in front of the target gap is considered as the reference vehicle for calculating the merging acceleration of the merging lane. Let the initial station of this vehicle be $S_X(t=0) = x$, and the initial station of the merging vehicle be: $S_R(t=0) = r$.

Following this adjustment, the time variation for the merging vehicle to accelerate uniformly is as follows:

$$t'_2 = \frac{v_0 - v_{R0}}{3.6 a'_r}. \quad (26)$$

The station location of vehicle X in front of the target gap at time $t_1 + t_2$ is

$$S_X(t_1 + t_2) = x + \frac{v_0}{3.6}(t_1 + t_2). \quad (27)$$

The station location of the merging vehicle at time $t_1 + t_2$ is

$$S'_R(t_1 + t_2) = \frac{(\frac{v_0}{3.6})^2 - (\frac{v_{R0}}{3.6})^2}{2a'_r}. \quad (28)$$

To ensure safe merging of the merging vehicle into the selected target gap, equation (29) should be satisfied.

$$S_X(t_1 + t_2) - L_v - L_{r-X} = S'_{ramp}(t_1 + t_2) \quad (29)$$

By substituting equations (24), (26), (27) and (28) into equation (29), the adjusted acceleration of the merging vehicle can be solved as

$$a'_r = -\frac{(v_0 - v_{R0})^2}{25.92(x + \frac{v_0|r|}{v_{R0}} - L_v - L_{r-X})}, \quad (30)$$

where $a'_r$ is the adjustment of the acceleration of vehicles on the acceleration lane, measured in meters per second squared (m/s²). $r$ is initial station of the ramp vehicle, measured in meters. $b$ is initial station of mainline vehicle B, measured in meters. $v_0$ is the traveling speed of the mainline vehicle, measured in kilometers per hour (km/h). $v_{R0}$ is the traveling speed of the ramp vehicle, measured in kilometers per hour (km/h). $L_v$ is vehicle length, taken as 5 meters, measured in meters. $L_{r-B}$ is safety distance to be maintained between the ramp vehicle and mainline vehicle during merging, measured in meters.

Therefore, the adjusted function representing the relationship between the station number and time for the entire travel process of the ramp vehicle is shown in equation (31).

$$S'_R(t) = \begin{cases} r + \frac{v_{R0}}{3.6}t, & 0 \leq t < t_1 \\ \frac{v_{R0}}{3.6}(t-t_1) + \frac{1}{2}a'_r(t-t_1)^2, & t_1 \leq t \leq t_1 + t'_2 \\ S'_2(t_1 + t_2) + \frac{v_0}{3.6}(t - t_1 - t_2), & t > t_1 + t'_2 \end{cases} \quad (31)$$

In summary, the gap between the mainline vehicle ramp vehicle and its preceding or following vehicle is chosen as the target gap for the ramp vehicle to merge into. After adjustment, the travel trajectories of the mainline vehicle and ramp vehicle are shown in Figure 6.

All mainline vehicles travel at a constant speed of $v_0$ without adjustments until the merging process is complete;

Ramp vehicles travel at a constant speed of $v_{R0}$ on the ramp until they reach the end of the ramp (station 0m). From the end of the ramp, they accelerate with the acceleration of $a'_r$ until they match the speed of the mainline vehicles. They merge onto the mainline at point $S'_{ramp}(t_1 + t_2)$, then travel at a constant speed of $v_0$ on the mainline.

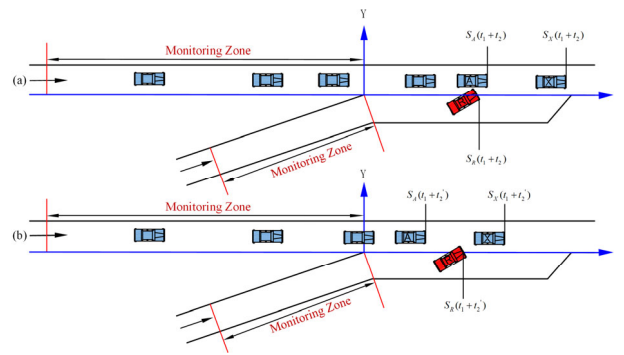

**FIGURE 6.** Vehicle Positions Before and After Adjustments. (a) Vehicle positions before adjustment, in which the ramp vehicle collides with the mainline vehicle; (b) Vehicle positions after adjustment, in which the ramp vehicle merges onto the mainline.

Furthermore, the target gap for the ramp vehicle to merge into is selected as the gap between the mainline vehicle and any vehicles other than its preceding and following vehicles, which is equal to or greater than the minimum merge gap for the ramp vehicle.

After selecting the gap, the acceleration after the adjustment is calculated using the method described above. After calculation, the velocity must satisfy the following conditions:

$$\frac{(\frac{v_0}{3.6})^2 - (\frac{v_{R0}}{3.6})^2}{2S_a} \leq a_r' \leq 6 \ m/s^2. \quad (32)$$

After selecting the gap, we calculate the merging point for the ramp vehicle after adjustment, ensuring that it satisfies the following conditions:

$$0 < S_R(t_1 + t_2) = \frac{v_{R0}t_2}{3.6} + \frac{1}{2}a_r t_2^2 < 200 \ m. \quad (33)$$

If the calculated acceleration for the ramp vehicle's travel and the merging point meet the requirements, then the function describing the relationship between the station number and time for the ramp vehicle after adjustment remains the same as equation (31).

By selecting a gap smaller than the minimum merge able gap $L_{\min s}$ as the target gap for the ramp vehicle to merge into, adjustments to the velocities of the vehicles before and after the target gap are necessary to accommodate the ramp vehicle's merge.

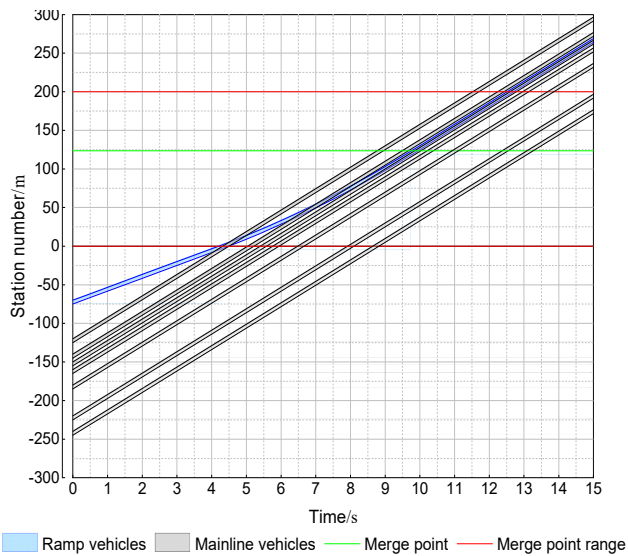

**FIGURE 7.** Pre-adjustment Spatiotemporal Diagram of Vehicle Collisions.

Before adjustment, the spatiotemporal diagram illustrating the variation in station numbers with respect to time for seven mainline vehicles and one ramp vehicle is depicted in Figure 7. A localized enlargement of this diagram is shown in Figure 8. In this depiction, the mainline vehicles travel at a speed of 100 km/h, whereas the ramp vehicle travels at 60 km/h on the ramp and accelerates at a rate of 2 m/s² on the acceleration lane until merging directly onto the mainline.

If either the calculated acceleration for the ramp vehicle's travel or the merging point does not meet the requirements, then we will not select the gap greater than or equal to the minimum merge able gap as the target gap. Instead, we choose a gap smaller than the minimum merge gap $L_{\min s}$, between the mainline vehicle conflicting with the ramp vehicle and its preceding vehicle, as the target gap for the ramp vehicle to merge into. Adjustments to the velocities of the mainline vehicles before the target gap are made to create sufficient space for the ramp vehicle to merge.

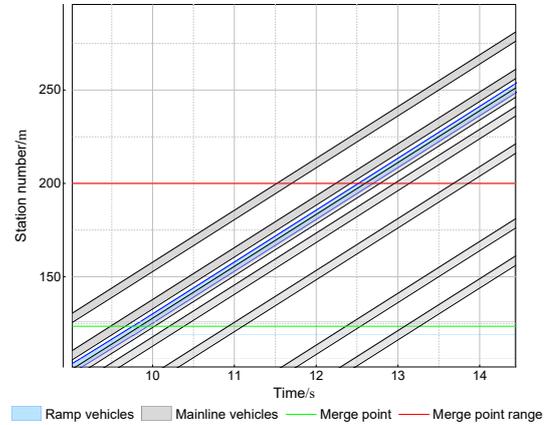

**FIGURE 8.** Localized Enlargement of the Pre-adjustment Spatiotemporal Diagram of Vehicle Collisions.

Let the initial station number of the mainline vehicle X before the selected target gap be: $S_X(t=0) = x$, and let the initial station number of the mainline vehicle Y after the selected target gap be: $S_Y(t=0) = y$. The initial number of stations the ramp vehicle was $S_R(t=0) = r$. The time required for the ramp vehicle to travel from its initial station to the end of the ramp is $t_1 = \frac{|r|}{v_{R0}/3.6}$. Therefore, at time $t = t_1$, the positions of mainline vehicles X and Y before and after the target gap, respectively, are shown in equation (34) and equation (35).

$$S_X(t = t_1) = x + \frac{v_0 t_1}{3.6} = x + \frac{v_0 |r|}{v_{R0}} \quad (34)$$

$$S_Y(t = t_1) = y + \frac{v_0 t_1}{3.6} = y + \frac{v_0 |r|}{v_{R0}} \quad (35)$$

When it is necessary to adjust the velocities of mainline vehicles before and after the target gap to ensure that there is no gap greater than or equal to the minimum merge gap on the mainline, the process involves creating space for the ramp vehicle to merge safely into the minimum merge gap. Subsequently, based on the relevant data, the feasibility of the ramp vehicle merging safely into the gap left by the mainline vehicles before and after the target gap is calculated. The specific approach for the entire process is as follows:

During the adjustment process to create the minimum merge gap for the ramp vehicle between the mainline vehicles before and after the target gap, there are three scenarios for the velocity changes in vehicles X and Y:

Scenario 1: Mainline vehicle Y continues to travel at its original speed, whereas vehicle X accelerates uniformly to create the minimum merge gap between the two vehicles for the ramp vehicles.

Scenario 2: Mainline vehicle X continues to travel at its original speed, whereas vehicle Y decelerates uniformly to create the minimum merge gap for the ramp vehicle between the two vehicles.
Scenario 3: Mainline vehicle X accelerates uniformly whereas vehicle Y decelerates uniformly until the minimum merge gap for the ramp vehicle is created between the two vehicles.

Scenario 1 only affected the vehicles before mainline vehicle X, and Scenario 2 only affected the vehicles after mainline vehicle Y. Scenario 3 caused the greatest disturbance to the mainline vehicles because both mainline vehicles X and Y needed to adjust their speeds. To ensure that vehicles X and Y maintain a safe distance from their preceding and following vehicles during the speed adjustment process, it may be necessary to adjust the vehicles before vehicle X or after vehicle Y. This adjustment could affect the vehicles within a certain range before mainline vehicle X and after mainline vehicle Y. Particularly when the spacing between mainline vehicles and their preceding or following vehicles is small, adjusting the speeds of both vehicles may require further adjustments to maintain the minimum safe distance between vehicles. Although Scenario 2 has a smaller impact range than Scenario 3, when the spacing between the vehicles after mainline vehicle Y is small, the deceleration of vehicle Y can significantly affect that vehicles follow it. Therefore, considering the above factors, scenario 1 was chosen as the adjustment plan for the mainline vehicle speeds. The specific adjustment plan for vehicle X is described in the following text.

The specific process of adjusting vehicle X is as follows:
The minimum distance that vehicle X needs to travel forward to create the minimum target merge gap is

$$\Delta L_X = L_{\min s} - L, \quad (36)$$

where $\Delta L_X$ is the distance that the vehicle must travel forward to create the minimum merge gap measured in meters. $L_{\min s}$ is the minimum safe gap that is required for the ramp vehicle to merge into the mainline measured in meters. $L$ is the selected target gap length is, $L = x - y - 5$, measured in meters.

During the process of adjusting the speed of vehicle X to create a mergeable gap for the ramp vehicle, vehicle Y continues to travel forward at a constant speed $v_0$. After time $t_2'$ has elapsed, a mergeable gap is formed between vehicles X and Y for the ramp vehicle. Throughout this process, vehicle Y travels a distance of $\frac{v_0 t_2'}{3.6}$ m. Vehicle X needs to travel an additional $\Delta L_X$ m compared to vehicle Y. Therefore, the distance that vehicle A needs to travel within this time is $(\frac{v_0 t_2'}{3.6} + \Delta L_X)$ m. The diagram of the positions of vehicles X and Y before and after creating the mergeable gap is shown in Figure 9.

Graph (a) illustrates the positions of various vehicles when the ramp vehicle reaches the end of the ramp; graph (b) represents the positions of various vehicles before the ramp vehicle merges into the mainline and graph (c) shows the positions of various vehicles after the ramp vehicle merges into the mainline.

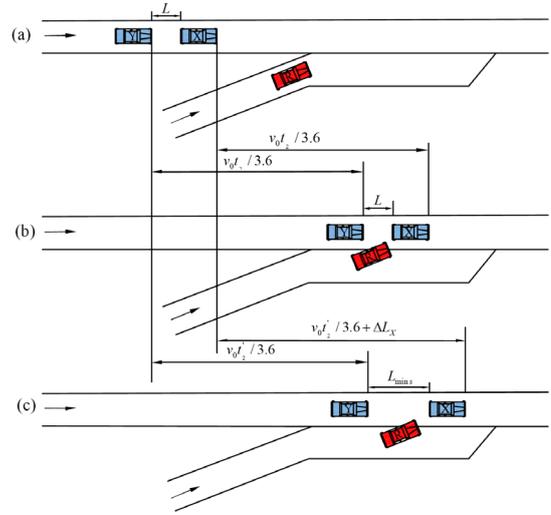

**FIGURE 9. Positions Before and After the Adjustment**

After the adjustment, and $t_2'$ s, the ramp vehicle accelerates from the marker at 0m to the merge point. The acceleration of the ramp vehicle in the acceleration lane after adjustment is

$$a_r' = \frac{v_0 - v_{R0}}{3.6 t_2'}. \quad (37)$$

The position of the merge point for the ramp vehicle at time $t_1 + t_2$ after the adjustment is

$$S_{mr}' = \frac{(\frac{v_0}{3.6})^2 - (\frac{v_{R0}}{3.6})^2}{2 a_r'} = \frac{(v_0 + v_{R0}) \cdot t_2'}{7.2}. \quad (38)$$

At the merge point, the relationship between the position of the ramp vehicle and that of the mainline vehicle X ahead of the target gap is

$$S_X = S_{mr}' + L_v + L_{r-X} - (\frac{v_0 t_2'}{3.6} + \Delta L_X). \quad (39)$$

To minimize the impact of the speed adjustments of mainline vehicle X on other vehicles on the mainline, ensuring that the speed of vehicle X remains the same as the original speed after leaving a sufficient safety gap, vehicle X accelerates uniformly for the first half of the entire speed adjustment process and decelerates uniformly for the second half with an acceleration equal to the opposite of the acceleration during acceleration, denoted as $a_{Xdec} = -a_{Xacc}$. Additionally, the absolute values of the acceleration during the first half and the second half of the process are equal, denoted as $a_{Xdec} = -a_{Xacc}$. Therefore, the time taken for acceleration $t_{Xacc}$ and deceleration $t_{Xdec}$ is the same, both equal to $\frac{t_2'}{2}$ s. The distance traveled during acceleration and deceleration is also the same, denoted as $S_{Xacc} = S_{Xdec} = \frac{S_X}{2} = \frac{\frac{v_0 t_2'}{3.6} + \Delta L_X}{2}$ m.

The relationship between distance and time during acceleration is

$$S_X = \frac{v_0}{3.6} \cdot t'_2 + \Delta L_X. \qquad (40)$$

The relationship between distance and acceleration during acceleration is

$$\frac{S_X}{2} = \frac{v_0}{3.6} \cdot \frac{t'_2}{2} + \frac{1}{2} a_{Xacc} (\frac{t'_2}{2})^2. \qquad (41)$$

Therefore, find the acceleration of the mainline vehicle X traveling with uniform acceleration is

$$a_{Xacc} = \frac{4 \Delta L_X}{(t'_2)^2}, \qquad (42)$$

where $a_{Xacc}$ is the acceleration of vehicle X when it is about to merge into the target gap before the exit ramp, measured in m/s². $t'_2$ is the total time vehicle X reserves before the target gap to merge into it, measured in meters. $S_X$ is the total distance that mainline vehicle X needs to travel throughout the entire deceleration process, measured in meters.

Due to $a_{Xdec} = -a_{Xacc}$, therefore

$$a_{Xdec} = -a_{Xacc} = -\frac{4 \Delta L_X}{(t'_2)^2}. \qquad (43)$$

The calculation formula for the total acceleration time and deceleration time of vehicle X adjusting its speed before the target gap, forming a period during which ramp vehicles can merge into the gap, is given in equation (44).

$$t_{Xdec} = t_{Xacc} = \frac{t'_2}{2} \qquad (44)$$

Speed of the mainline vehicle after uniform acceleration is

$$v_t = \frac{v_0}{3.6} + a_{Xacc} \cdot t_{Xacc}. \qquad (45)$$

Using the collaborative control method described above, the acceleration of the ramp vehicles was calculated. In addition, the vehicles that would collide after the merging point are adjusted. The spatiotemporal diagram of the adjusted vehicles is shown in Figure 10, with a magnified view provided in Figure 11.

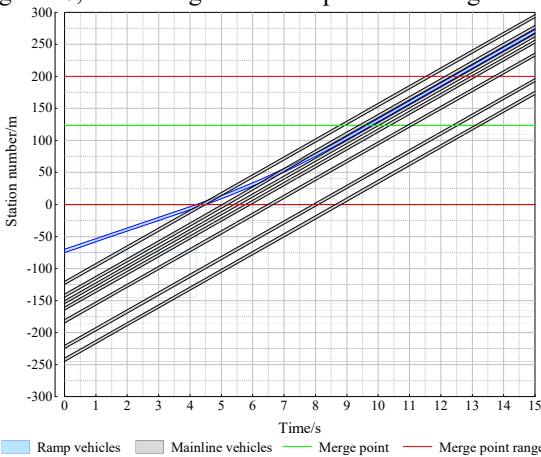

**FIGURE 10.** Spatiotemporal Diagram of the Adjusted Vehicles.

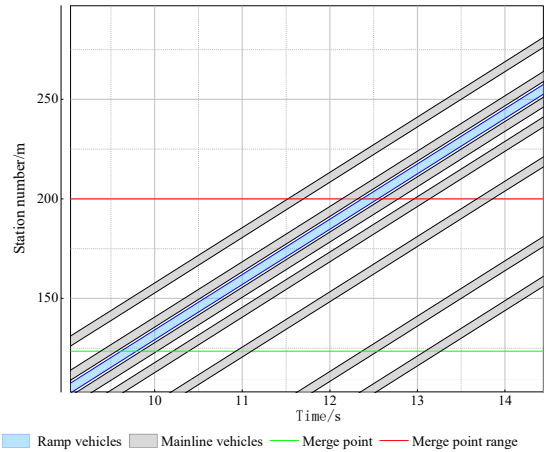

**FIGURE 11.** Zoomed-in Spatiotemporal Diagram of Adjusted Vehicles.

## V. Simulation Experiments
### A. SIMULATION FRAMEWORK
Utilizing Python3.10 and SUMO1.19, we built a simulation framework as shown in Figure 12. In this framework, Python is mainly used to implement major algorithm modules such as calculation processing and trajectory generation. The established road network is shown in Figure 13.

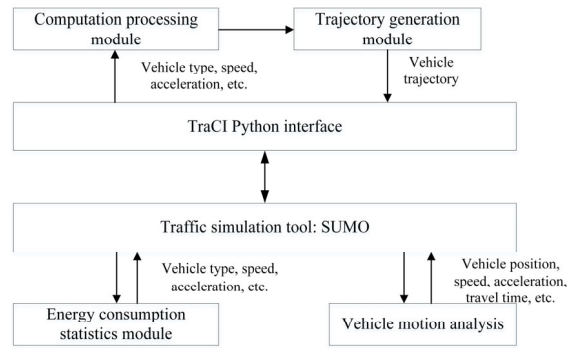

**FIGURE 12.** Simulation Framework.

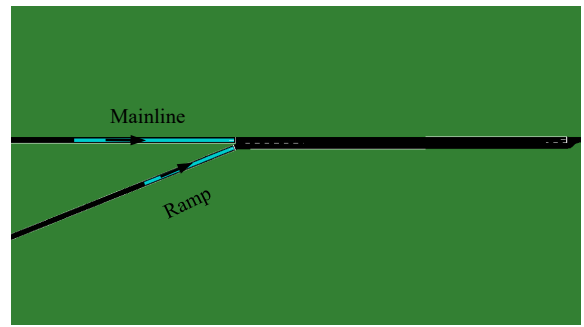

**FIGURE 13.** Road Network.

### B. SIMULATION SCENE AND PARAMETER SETTINGS
#### 1) SIMULATION SCENE
The simulation scene selected in this section is mainly set in the outermost lane of the mainline expressway in a connected environment and at the single-lane entrance section of the ramp. It was assumed that the maximum speed of the vehicles on the ramp was 17 m/s, while the speed of vehicles on the mainline varied between 15 and 20 m/s. The acceleration lane length is set to 200 m.

Simulation was conducted with mainline traffic volumes of 800 veh/h/lane, 1200 veh/h/lane, and 1800 veh/h/lane, and ramp traffic volumes of 200 veh/h/lane, 300 veh/h/lane, and 500 veh/h/lane. The mainline and ramp traffic volumes are randomly paired to form the simulation scenarios.

All connected autonomous vehicles are equipped with vehicle-infrastructure cooperative devices, providing functions such as vehicle positioning and real-time communication with the infrastructure. All vehicles in the merging area of the expressway ramp obey the control and drive according to the planned trajectories.

The key to the ramp vehicles merging into the mainline at the ramp merging area on the expressway is to find a suitable merging gap. The mainline priority strategy means looking for the target gap on the main line and allowing the ramp vehicles to seek opportunities to merge into the main line. On the other hand, the ramp priority strategy determines the merging position of the ramp vehicles, enabling the vehicles before and after the target gap on the main line to travel freely or adjusting the main line vehicles so that the ramp vehicles can reach the designated position and then merge into the main line just in time.

#### 2) MODEL AND PARAMETER
Because previous studies have not proposed pre-planned trajectories for vehicles under mainline priority and ramp priority cooperative control methods, the two cooperative control methods proposed in this paper are selected for comparison with no cooperative control. The no cooperative control method uses the Krauss car-following model and LC2013 lane-changing model in SUMO. Therefore, the simulation conducted in the SUMO software using the Krauss model as the car-following model and the LC2013 model as the lane-changing decision model is referred to as the Krauss/LC2013 model in the subsequent text.

### C. SIMULATION RESULTS
#### 1) AVERAGE DELAY TIME
The average delay is a crucial traffic metric, particularly when evaluating the effectiveness of expressway ramp merging, as it provides key insights into traffic flow and efficiency. This reflects traffic congestion, merging efficiency, traffic safety, and the effectiveness of cooperative control.

In this study, simulations were conducted with mainline traffic volumes of 800, 1200 and 1800 veh/h/lane, and ramp traffic volumes of 200 veh/h/lane, 300 veh/h/lane, and 500 veh/h/lane. Three strategies were simulated: mainline priority, ramp priority, and Krauss/LC2013. This resulted in nine different traffic flow scenarios with the corresponding average delay times, as shown in TABLE 1.

The average delay for mainline vehicles is the additional travel time compared with free-flow conditions. Similarly, the average delay for ramp vehicles is the extra travel time compared to free-flow conditions. For mainline vehicles, the free-flow travel time over 800m in the ramp merging area is 40s, while for ramp vehicles, it was 25s over 400m in the merging area. Improvement rate of average delay ($I_r$) of the mainline priority and improvement rate of average delay ($I_p$) of the ramp priority are

$$I_r = (1 - \frac{D_m}{D_k}) \cdot 100\% \quad (46)$$

and

$$I_p = (1 - \frac{D_r}{D_k}) \cdot 100\% , \quad (47)$$

where $D_m$ is the ramp average delay time under the mainline priority, measured in meters; $D_k$ is the ramp average delay time under the Krauss/LC2013 model, measured in meters. $D_r$ is the ramp average delay time under the mainline priority, measured in meters.

**TABLE 1. Average Delay for Mainline Traffic.**

| Mainline traffic volume (veh/h/lane) | Ramp traffic volume (veh/h/lane) | Mainline average delay time | | | | |
|---|---|---|---|---|---|---|
| | | $D_k$ (s) | $D_m$ (s) | $I_r$ (%) | $D_r$ (s) | $I_p$ (%) |
| 800 | 200 | 4.44 | 0.23 | 94.82 | 0.61 | 86.26 |
| 800 | 300 | 5.01 | 0.38 | 92.42 | 1.10 | 78.04 |
| 800 | 500 | 5.99 | 0.45 | 92.49 | 0.97 | 83.81 |
| 1200 | 200 | 7.60 | 0.26 | 96.58 | 0.54 | 92.89 |
| 1200 | 300 | 8.55 | 0.57 | 93.33 | 1.19 | 86.08 |
| 1200 | 500 | 9.45 | 0.54 | 94.29 | 1.31 | 86.14 |
| 1800 | 200 | 15.28 | 0.54 | 96.47 | 0.73 | 95.22 |
| 1800 | 300 | 18.04 | 0.65 | 96.40 | 1.24 | 93.13 |
| 1800 | 500 | 26.98 | 0.68 | 97.48 | 1.35 | 95.00 |

From TABLE 1, it can be observed that under the same traffic conditions, the average delay time of the mainline vehicles under the mainline priority strategy is always lower than that under the ramp priority strategy. Additionally, under the same traffic volume, the improvement rate of the mainline average delay time under the mainline priority strategy, compared to the Krauss/LC2013 model used in SUMO, was consistently higher than that under the ramp priority strategy. In the simulated traffic volumes, the improvement rate of the mainline vehicle average delay under the mainline priority strategy can reach up to 97.48%, whereas under the ramp priority strategy, it reaches a maximum of 95%. When the mainline traffic volume is fixed, in most cases, with an increase in ramp traffic volume, the average delay of mainline vehicles under both mainline priority and ramp priority strategies increases, while the improvement rate decreases.

**TABLE 2. Average Delay for Ramp Traffic.**

| Mainline traffic volume (veh/h/lane) | Ramp traffic volume (veh/h/lane) | Ramp average delay time | | | | |
|---|---|---|---|---|---|---|
| | | $D_k$ (s) | $D_m$ (s) | $I_r$ (%) | $D_r$ (s) | $I_p$ (%) |
| 800 | 200 | 7.35 | 0.15 | 97.96 | 0.46 | 93.74 |
| 800 | 300 | 8.62 | 0.31 | 96.40 | 0.43 | 95.01 |
| 800 | 500 | 9.65 | 0.33 | 96.58 | 0.37 | 96.17 |
| 1200 | 200 | 8.54 | 0.18 | 97.89 | 0.45 | 94.73 |

| Mainline traffic volume (veh/h/lane) | Ramp traffic volume (veh/h/lane) | Ramp average delay time | | | | |
|---|---|---|---|---|---|---|
| | | $D_k$(s) | $D_m$(s) | $I_r$(%) | $D_r$(s) | $I_p$(%) |
| 1200 | 300 | 9.66 | 0.34 | 96.48 | 0.44 | 95.45 |
| 1200 | 500 | 12.29 | 0.44 | 96.42 | 0.67 | 94.55 |
| 1800 | 200 | 11.47 | 0.55 | 95.20 | 0.71 | 93.81 |
| 1800 | 300 | 16.43 | 0.66 | 95.98 | 0.86 | 94.77 |
| 1800 | 500 | 20.42 | 0.59 | 97.11 | 0.85 | 95.84 |

From TABLE 2, the following conclusions can be drawn regarding the average delay time for different ramp traffic volumes: Under all traffic flow conditions, the average delay time for ramps under the mainline priority strategy is always lower than that under the Krauss/LC2013 model and ramp priority strategy. Additionally, the improvement rate of the mainline priority strategy is higher than that of the ramp priority strategy for different traffic flow volumes, indicating a better performance in reducing ramp vehicle delays. Although the average delay time for ramps under the ramp priority strategy is also lower than that under the Krauss/LC2013 model, its improvement rate is consistently lower than that of the mainline priority strategy.

Under the conditions of high mainline traffic volume and low ramp traffic volume, the mainline priority strategy can significantly reduce delay time, showing the most pronounced effect. Therefore, the mainline priority strategy performs better than the ramp priority strategy in reducing ramp vehicle delays, particularly under conditions of high mainline traffic volume and low ramp traffic volume.

In the simulated traffic volumes, the improvement rate of the ramp vehicle average delay under the mainline priority strategy can reach up to 97.96%, whereas under the ramp priority strategy, it reaches a maximum of 96.17%.

The mainline average delay indicates:
(a) Mainline traffic flow status: The average delay on the mainline reflects the smoothness of main road traffic. Higher delays indicate that mainline traffic may approach or exceed the road capacity, resulting in speed reduction and increased travel time.
(b) Merge impact: The increased average delay on the mainline may be caused by ramp merging, especially if the merging design is improper or the merging traffic volume is too high, thereby increasing the delays.

### 2) AVERAGE SPEED

The average speed is a key indicator for assessing the efficiency and safety of expressway ramp mergers. A higher average speed typically indicates smooth merging onto the main road, indicating good traffic flow in an area with no significant congestion. The stability of average speed indicates the predictability and stability of traffic flow, which is crucial for reducing the risk of accidents caused by sudden braking or acceleration.

When analyzing the average speed of expressway ramp merging areas, three scenarios are selected: the maximum mainline flow rate (1800 veh/h/lane) with ramp flow rates of 200, 300, and 500 veh/h/lane. If a particular cooperative control method performs better with higher mainline traffic flow, it may perform better with a lower mainline traffic flow. The average mainline speed is shown in Figure 14, and the average ramp speed is shown in Figure 15. In both figures:
(a) represents the scenario with a mainline flow rate of 1800 veh/h/lane and ramp flow rate of 200 veh/h/lane;
(b) represents the scenario with a mainline flow rate of 1800 veh/h/lane and ramp flow rate of 300 veh/h/lane;
(c) represents the scenario with a mainline flow rate of 1800 veh/h/lane and ramp flow rate of 500 veh/h/lane.

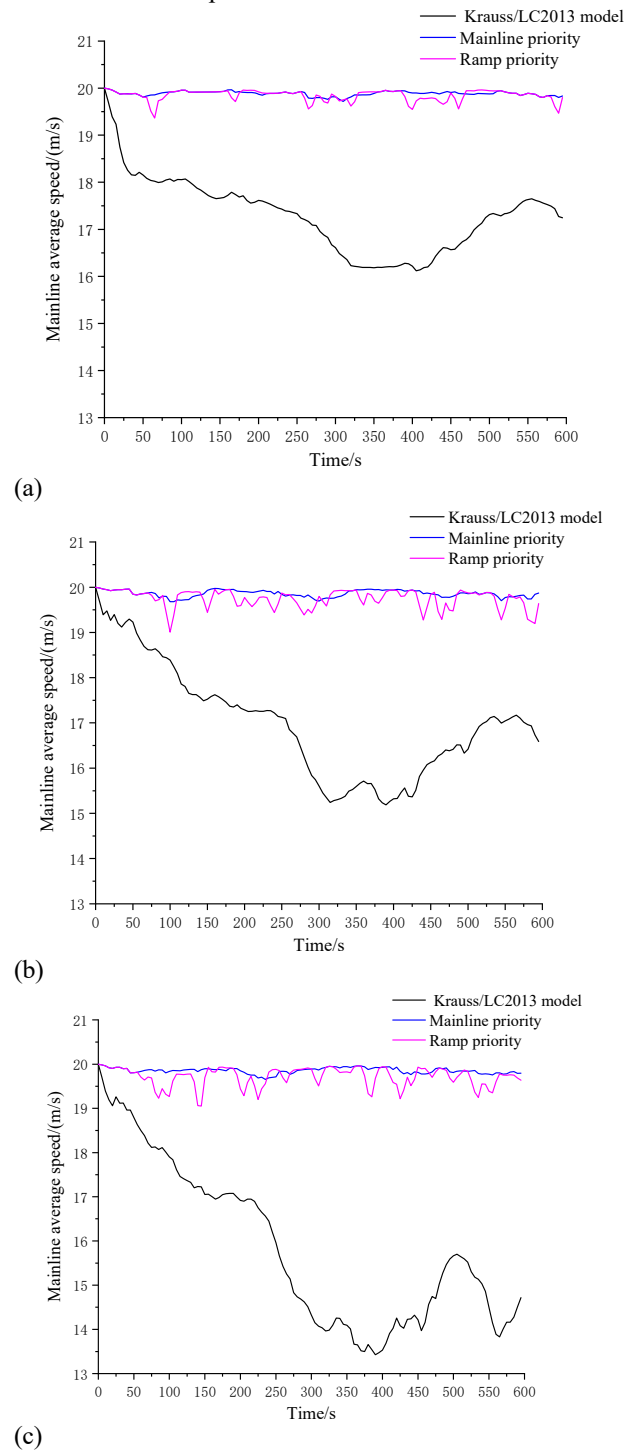

**FIGURE 14.** Mainline Average Speed.

From Figure 14, it can be observed that in expressway ramp merging under three different traffic flow conditions, the Krauss/LC2013 model resulted in the lowest and most fluctuating average speed on the mainline. Therefore, the mainline priority control strategy demonstrated a more balanced performance in terms of reducing speed fluctuations and maintaining traffic stability.

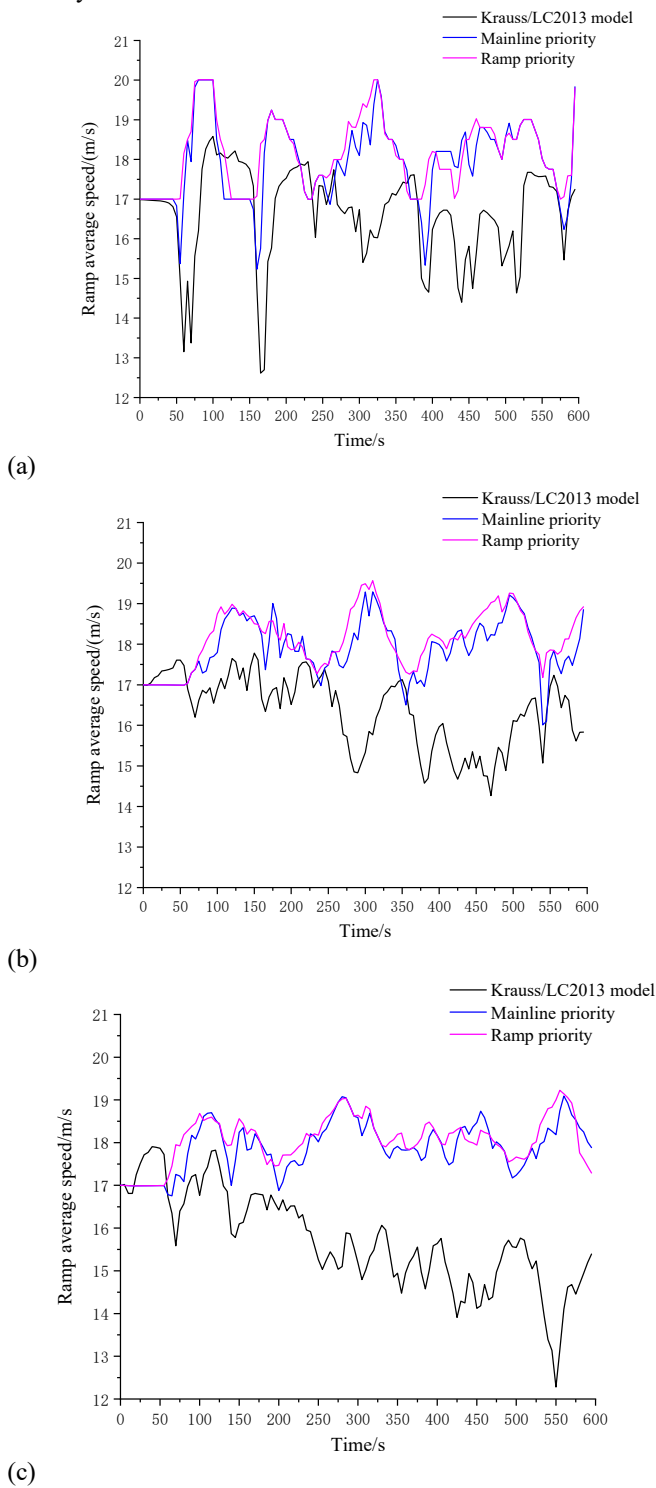

(a)

(b)

(c)

**FIGURE 15.** Ramp Average Speed.

From Figure 15, it can be observed that in expressway ramp merging under three different traffic flow conditions, we found that the Krauss/LC2013 model resulted in the lowest average speed and the largest speed fluctuations for ramp vehicles. This suggests that this strategy may have failed to effectively coordinate the traffic between the ramp and mainline, leading to difficulties in ramp merging and traffic congestion. In contrast, both the mainline priority and ramp priority strategies exhibited similar and relatively stable average speeds and fluctuation ranges, indicating that these two methods are more effective in managing traffic flow and can better balance the needs of the ramp and the mainline, thereby reducing congestion and improving traffic flow efficiency and safety.

In summary, in expressway ramp merging under three different traffic flow conditions, the Krauss/LC2013 model showed the lowest average speed and the largest speed fluctuations for ramp vehicles, indicating that this strategy was not effective enough to coordinate traffic between the ramp and the mainline, leading to merging difficulties and increased traffic congestion. While the mainline priority and ramp priority strategies exhibited similar higher average speeds, suggesting their effectiveness in cooperative control, the larger speed fluctuations in the ramp priority strategy indicated that they might cause frequent speed adjustments for mainline traffic, affecting driving comfort and safety. In contrast, the mainline priority strategy optimized traffic stability with smaller speed fluctuations, achieving the best balance between efficiency and safety.

### 3) TOTAL FUEL CONSUMPTION

In this study, when the acceleration was negative, $f_{accel}$ was automatically set to zero. In the fuel consumption model, setting $f_{accel}$ to zero when it is negative is mainly because in this case, the vehicle decelerating or driving downhill, and its fuel consumption is no longer directly driven by acceleration. Firstly, fuel consumption is mainly influenced by the additional energy demand during acceleration, which decreases during deceleration, thus, the related fuel consumption can be considered to be zero. Secondly, modern vehicles are often equipped with energy recovery systems that can recover kinetic energy during deceleration, reduce the overall energy consumption and further reducing the necessity of considering $f_{accel}$ in this state. In addition, simplifying the model is an important consideration in the modeling process; not calculating $f_{accel}$ during deceleration can reduce model complexity without significantly affecting the overall accuracy of the model. Finally, during vehicle deceleration, fuel consumption is influenced more by factors such as engine braking, air resistance, and rolling resistance, rather than the deceleration action itself. Therefore, this approach reflects both physical reality and provides convenience and accuracy in the model development.

Based on the engine operating conditions, fuel consumption models can generally be classified into steady-state and transient models. In this study, we have adopted a simplified fuel consumption model[22] and plan to further optimize and refine the model in future research.

Where $p_0$, $p_1$, $p_2$, $p_3$, $q_0$, $q_1$ and $q_2$ are coefficients of the polynomial and are constants. The values of each parameter in the energy consumption model are shown in TABLE 3, and the graph of the fuel consumption model with these parameter values is illustrated in Figure 16.

TABLE 3. Parameter Values of the Energy Consumption Model.

| Parameters | Parameter meanings | Values | Units |
|---|---|---|---|
| $M_v$ | Vehicle mass | 1200 | Kg |
| $C_D$ | Drag coefficient | 0.32 | \ |
| $\rho_\alpha$ | Air density | 1.184 | Kg/m$^3$ |
| $A_f$ | Vehicle frontal area | 2.5 | m$^2$ |
| $\mu$ | Viscosity coefficient | 0.015 | \ |
| $p_0$ | Polynomial coefficient | 0.1569 | \ |
| $p_1$ | Polynomial coefficient | 0.0245 | \ |
| $p_2$ | Polynomial coefficient | $-7.415 \times 10^{-4}$ | \ |
| $p_3$ | Polynomial coefficient | $5.975 \times 10^{-5}$ | \ |
| $q_0$ | Polynomial coefficient | 0.07224 | \ |
| $q_1$ | Polynomial coefficient | 0.09681 | \ |
| $q_2$ | Polynomial coefficient | $1.075 \times 10^{-3}$ | \ |

The speed of the vehicle is represented on the x-axis, acceleration on the y-axis, and fuel consumption on the z-axis. The velocity ranges from 0 to 30 m/s, and the acceleration ranges from 0 to 5 m/s$^2$. The functional relationship between fuel consumption, velocity, and acceleration is

$$f_{fuel} = 0.1596 + 0.0245v - 7.145 \times 10^{-4}v^2 + 5.975 \times 10^{-5}v^3 + a(0.07224 + 0.09681v + 1.075 \times 10^{-3}v^2). \quad (48)$$

The three-dimensional plot illustrating the relationship between fuel consumption, velocity, and acceleration, based on the aforementioned functional relationship, is shown in Figure 16.

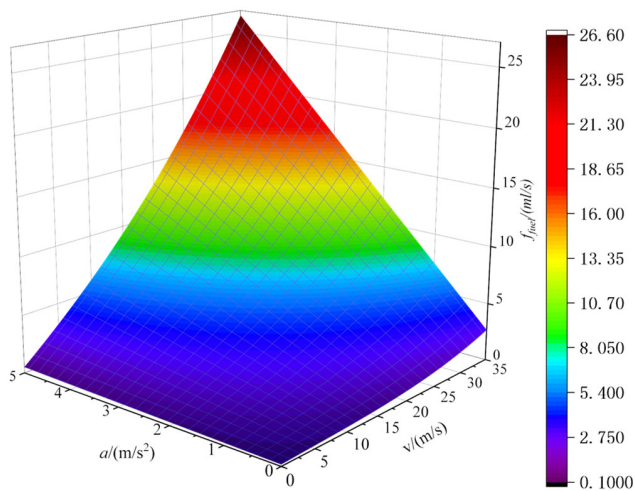

FIGURE 16. Fuel Consumption Model.

In summary, the mainline priority strategy performs better than the merging lane priority strategy in reducing total fuel consumption. Under the same traffic flow conditions, the maximum improvement rate of the total fuel consumption for the mainline priority strategy reached 6.01%, whereas that for the merging lane priority strategy reached 4.88%.

TABLE 4. Fuel Consumption by Traffic Volume & Control Methods.

| Mainline traffic volume (veh/h/lane) | Ramp traffic volume (veh/h/lane) | Krauss /LC2013 /(L) | Mainline priority /(L) | $I_r$ /% | Ramp priority /(L) | $I_p$ /% |
|---|---|---|---|---|---|---|
| 800 | 200 | 15.18 | 15.04 | 0.92 | 15.13 | 0.33 |
| 800 | 300 | 16.93 | 16.47 | 2.72 | 16.3 | 3.72 |
| 800 | 500 | 19.65 | 18.91 | 3.77 | 19.17 | 2.44 |
| 1200 | 200 | 20.93 | 20.62 | 1.48 | 20.59 | 1.62 |
| 1200 | 300 | 22.36 | 21.77 | 2.64 | 21.48 | 3.94 |
| 1200 | 500 | 25.02 | 24.27 | 3.00 | 23.8 | 4.88 |
| 1800 | 200 | 28.56 | 28.49 | 0.25 | 27.49 | 3.75 |
| 1800 | 300 | 29.91 | 29.64 | 0.90 | 28.86 | 3.51 |
| 1800 | 500 | 34.76 | 32.67 | 6.01 | 32.95 | 5.21 |

From TABLE 4, the following conclusions can be drawn:

1) The mainline priority strategy consistently results in lower total fuel consumption than the Krauss/LC2013 model across all traffic volumes, indicating its effectiveness in reducing fuel consumption. Moreover, its improvement rate was higher than that of the ramp priority strategy across different traffic volumes.

2) The ramp priority strategy also exhibits lower total fuel consumption than the Krauss/LC2013 model. Although its improvement rate is not as high as that of the mainline priority strategy, it still demonstrates some optimization effect.

3) With the increase in mainline and ramp traffic volume, the total fuel consumption under the Krauss/LC2013 model significantly increases, indicating that higher traffic pressure leads to higher fuel consumption.

4) Under a constant mainline traffic volume, the total fuel consumption gradually increases with an increase in ramp traffic volume for both the mainline priority and ramp priority strategies, indicating that a higher ramp traffic volume leads to increased fuel consumption. In particular, under the conditions of mainline traffic volume at 1800 veh/h/lane and ramp traffic volume at 500 veh/h/lane, the mainline priority strategy achieves the highest improvement rate of 6.10%, while the ramp priority strategy has an improvement rate of 5.21%. This highlights the significant optimization effect of the mainline priority strategy under high mainline and ramp traffic volumes.

## VI. Discussion

In this study, a comprehensive approach is proposed to enhancing expressway ramp merge safety and efficiency through spatiotemporal cooperative control. The results obtained from our research have several significant implications and open up new avenues for further exploration.

Firstly, the proposed method for calculating the safe distance between vehicles in the vehicle-road depth cooperation scenario provides a more accurate and practical foundation for ensuring safe driving distance. By considering factors such as GPS positioning errors, clock synchronization errors, and speed differences, our approach addresses the real-world uncertainties that exist in the driving environment. This is in contrast to previous studies that may have overlooked or inadequately considered these factors. For example, in some existing research, the influence of positioning errors was either not incorporated or was overly simplified. Our method provides a more refined way to determine the safe distance, which can potentially reduce the risk of collisions caused by inaccurate distance estimations.

Secondly, the quantitative evaluation model of vehicle conflict risk based on collision acceleration and urgent acceleration is a novel contribution. This model facilitates a deeper understanding of the nature of vehicle conflicts. By combining these two key parameters, we can better assess the potential harm and urgency of collisions. Compared to other risk assessment models found in the literature, which may depend solely on speed variances or basic time-based metrics, our model provides a more comprehensive and precise representation of the actual collision risk. This allows for the development of more targeted and effective control strategies.

The mainline priority coordinated control method proposed in this paper also demonstrated its superiority. In the simulation experiments, compared with traditional following and lane-changing strategies, the mainline priority and ramp priority methods can reduce the average delay time by 97.96% and 96.17%, respectively and decrease fuel consumption by 6.01% and 4.88%, respectively. This indicates that our method can effectively balance the traffic flow between the mainline and the ramp, reducing congestion and improving overall traffic efficiency. For instance, the significant reduction in average delay time shows that vehicles can merge more smoothly, minimizing disruptions to the traffic stream. The enhanced fuel consumption efficiency also holds significant implications for environmental protection and energy conservation within the transportation sector.

Mainline priority coordinated control methods are proposed to pre-plan vehicle trajectories. Firstly, the safe distance between vehicles, as calculated using the proposed formula, is determined. Then, the conflict level of merge-lane vehicles is calculated to identify mainline vehicles that may conflict. Subsequently, the target gap for merge-lane vehicles is determined, and the corresponding coordinated control methods are established based on the selected target gap. Accordingly, vehicle trajectories are pre-planned and vehicles can safely merge based on the planned trajectories.

However, it is important to acknowledge the limitations of our study. One of the main limitations is the assumption of certain ideal conditions in the simulation. For instance, the traffic flow patterns are relatively simplified, and the effects of extreme weather conditions or sudden traffic incidents were not considered. Future research could expand the model to include these factors, making it more applicable in real-world scenarios. Additionally, the parameters used in the fuel consumption model were based on existing data and assumptions, and there may be opportunities for further refinement by collecting more detailed vehicle performance data.

Another aspect worthy of further investigation is the scalability of the proposed methods. With the continuous increase in the number of vehicles on the expressway, it is essential to ensure the effectiveness of the control strategies. Future studies could explore how the algorithms can be adapted to handle higher traffic densities and more complex traffic compositions.

In conclusion, while our research has made significant progress in improving expressway ramp merge safety and efficiency, there are still areas that require further research and improvement. By addressing the limitations and exploring the potential extensions discussed above, we can continue to enhance the performance of our methods and contribute to the development of more intelligent and efficient transportation systems.

## VII. CONCLUSION

This research has made substantial contributions to the field of expressway ramp merge control by introducing an innovative spatiotemporal cooperative control method. Through a series of further analyses and comprehensive simulations, we have achieved significant advancements in enhancing both the safety and efficiency of ramp merging processes. The approach we proposed, which centers around the calculation of safe distances under diverse spatiotemporal conditions and the development of a sophisticated vehicle conflict risk evaluation model, has provided a solid theoretical and practical foundation. By precisely quantifying critical factors such as collision acceleration and emergency acceleration, we have been able to more accurately assess potential risks and take proactive measures to mitigate them. This represents a crucial step forward compared to previous studies that often lacked such detailed and comprehensive risk assessment mechanisms. The mainline priority coordinated control method, a key component of our research, has demonstrated remarkable effectiveness. In extensive simulation scenarios with varying traffic volumes and conditions, it has consistently outperformed traditional methods. The significant reduction in average delay time, by up to 97.96%, clearly indicates that vehicles can merge into the mainline more smoothly and expeditiously, effectively alleviating traffic congestion at the ramp merge areas. Additionally, the notable decrease in fuel consumption, reaching 6.01%, not only brings economic benefits but also has positive implications for environmental sustainability. Nevertheless, we are fully aware of the existing limitations. For example, certain parameters, including vehicle output power, road gradient, and road surface friction coefficient, were not incorporated into the current model. These factors can potentially influence the performance of the control method in real-world applications. Future research will focus on integrating these parameters to further refine and enhance the accuracy and applicability of our approach.

In conclusion, this study offers valuable insights and practical solutions for the development of intelligent transportation systems. By continuing to address the identified limitations and explore new research directions, this work may inspire further advancements in this critical area of transportation research and lead to safer and more efficient expressway traffic.

## ACKNOWLEDGEMENT
The authors thank all the anonymous reviewers for their insightful comments.

## REFERENCES

[1] J. Chen, Y. Zhou, and E. Chung, "An integrated approach to optimal merging sequence generation and trajectory planning of connected automated vehicles for freeway on-ramp merging sections," *IEEE Transactions on Intelligent Transportation Systems,* vol. 25, no. 2, pp. 1897-1912, 2024.

[2] S. Li, C. Wei, and Y. Wang, "A ramp merging strategy for automated vehicles considering vehicle longitudinal and latitudinal dynamics, " *in 2020 IEEE 5th International Conference on Intelligent Transportation Engineering (ICITE)*, pp. 441-445.

[3] J. Zhu, L. Wang, I. Tasic, and X. Qu, "Improving freeway merging efficiency via flow-level coordination of connected and autonomous vehicles," *IEEE Transactions on Intelligent Transportation Systems,* vol. 25, no. 7, pp. 6703-6715, 2024.

[4] J. Q. Liu, W. Z. Zhao, C. Y. Wang, Y. N. Zhou, Z. Y. Zhang, and Y. J. Qin, "Interactive on-ramp merging decision-making with motion prediction considering merging motivation," *Proceedings of the Institution of Mechanical Engineers Part D-journal of Automobile Engineering*, Jul. 2023.

[5] H. G. Min, Y. K. Fang, R. M. Wang, X. C. Li, Z. G. Xu, and X. M. Zhao, "A novel on-ramp merging strategy for connected and automated vehicles based on game theory," *Journal of Advanced Transportation,* vol. 2020, Jul.2020.

[6] C. Wei, Y. He, H. Tian, and Y. Lv, "Game theoretic merging behavior control for autonomous vehicle at highway on-ramp," *IEEE Transactions on Intelligent Transportation Systems,* vol. 23, no. 11, pp. 21127-21136, 2022.

[7] S. C. Jing, F. Hui, X. M. Zhao, J. Rios-Torres, and A. J. Khattak, "Integrated longitudinal and lateral hierarchical control of cooperative merging of connected and automated vehicles at on-ramps," *IEEE Transactions on Intelligent Transportation Systems,* vol. 23, no. 12, pp. 24248-24262, Dec. 2022.

[8] Z. Hu, J. Huang, Z. Yang, and Z. Zhong, "Embedding robust constraint-following control in cooperative on-ramp merging," *IEEE Transactions on Vehicular Technology,* vol. 70, no. 1, pp. 133-145, 2021.

[9] D. Li, H. Pan, Y. Xiao, B. Li, L. Chen, H. Li, and H. Lyu, "Social-aware decision algorithm for on-ramp merging based on level-k gaming, " *in 2022 IEEE 18th International Conference on Automation Science and Engineering (CASE)*, pp. 1753-1758.

[10] Y. J. Xue, X. K. Zhang, Z. Y. Cui, B. Yu, and K. Gao, "A platoon-based cooperative optimal control for connected autonomous vehicles at highway on-ramps under heavy traffic," *Transportation research part c-emerging technologies,* vol. 150, May. 2023.

[11] T. Y. Chen, M. Wang, S. Y. Gong, Y. Zhou, and B. Ran, "Connected and automated vehicle distributed control for on-ramp merging scenario: A virtual rotation approach," *Transportation Research Part C-Emerging Technologies,* vol. 133, Dec.2021.

[12] S. Zhou, W. Zhuang, G. Yin, H. Liu, and C. Qiu, "Cooperative on-ramp merging control of connected and automated vehicles: Distributed multi-agent deep reinforcement learning approach," *in 2022 IEEE 25th International Conference on Intelligent Transportation Systems (ITSC)*, pp. 402-408.

[13] X. L. Luo, X. F. Li, M. R. R. Shaon, and Y. X. Zhang, "Multi-lane-merging strategy for connected automated vehicles on freeway ramps," *Transportmetrica B-Transport Dynamics,* vol. 11, no. 1, pp. 127-145, Dec. 2023.

[14] H. G. Min, Y. K. Fang, X. Wu, G. Y. Wu, and X. M. Zhao, "On-ramp merging strategy for connected and automated vehicles based on complete information static game," *Journal of Traffic and Transportation Engineering-English Edition,* vol. 8, no. 4, pp. 582-595, Aug.2021.

[15] Z. B. Gao, Z. Z. Wu, W. Hao, K. K. Long, Y. J. Byon, and K. J. Long, "Optimal trajectory planning of connected and automated vehicles at on-ramp merging area," *IEEE Transactions on Intelligent Transportation Systems,* vol. 23, no. 8, pp. 12675-12687, Aug. 2022.

[16] Z. J. Wang, A. Cook, Y. L. Shao, G. H. Xu, J. F. Chen, and Ieee, "Cooperative merging speed planning: A vehicle-dynamics-free method, " *presented at the 2023 IEEE Intelligent Vehicles Symposium*, IV, 2023.

[17] W. H. Chen, G. Ren, Q. Cao, J. H. Song, Y. K. Liu, and C. Y. Dong, "A game-theory-based approach to modeling lane-changing interactions on highway on-ramps: Considering the bounded rationality of drivers," *Mathematics,* vol. 11, no. 2, Jan.2023.

[18] G. Bagwe, J. Li, X. Deng, X. Yuan, and L. Zhang, "Poster: Reliable on-ramp merging via multimodal reinforcement learning, " *in 2022 IEEE/ACM 7th Symposium on Edge Computing (SEC)*, pp. 313-315.

[19] Z. E. Kherroubi, S. Aknine, and R. Bacha, "Novel decision-making strategy for connected and autonomous vehicles in highway on-ramp merging," *IEEE Transactions on Intelligent Transportation Systems,* vol. 23, no. 8, pp. 12490-12502, Aug.2022.

[20] J. Shi, Y. Luo, P. Li, Y. Hu, and K. Li, "An optimized scheduling method with dynamic conflict graph for connected and automated vehicles at multi-lane on-ramp areas, " *in 2023 IEEE Intelligent Vehicles Symposium (IV)*, pp. 1-7.

[21] J. Shi, K. Li, C. Chen, W. Kong, and Y. Luo, "Cooperative merging strategy in mixed traffic based on optimal final-state phase diagram with flexible highway merging points," *IEEE Transactions on Intelligent Transportation Systems,* vol. 24, no. 10, pp. 11185-11197, 2023.

[22] J. Rios-Torres, and A. A. Malikopoulos, "Automated and cooperative vehicle merging at highway on-ramps," *IEEE Transactions on Intelligent Transportation Systems,* vol. 18, no. 4, pp. 780-789, Apr.2017.

[23] Z. X. Tang, H. Zhu, X. Zhang, M. Iryo-Asano, and H. Nakamura, "A novel hierarchical cooperative merging control model of connected and automated vehicles featuring flexible merging positions in system optimization," *Transportation Research Part C-Emerging Technologies,* vol. 138, May.2022.

[24] N. Chen, B. v. Arem, and M. Wang, "Hierarchical optimal maneuver planning and trajectory control at on-ramps with multiple mainstream lanes," *IEEE Transactions on Intelligent Transportation Systems,* vol. 23, no. 10, pp. 18889-18902, 2022.

[25] J. X. Wu, Y. B. Wang, Z. Zhang, Y. Q. Wen, L. X. Zhong, and P. J. Zheng, "A cooperative merging control method for freeway ramps in connected and autonomous driving," *Sustainability,* vol. 14, no. 18, Sep.2022.

[26] S. D. Kumaravel, A. A. Malikopoulos, and R. Ayyagari, "Decentralized cooperative merging of platoons of connected and automated vehicles at highway on-ramps, " *in 2021 American Control Conference (ACC)*, pp. 2055-2060.

[27] C. T. Bian, G. D. Yin, L. W. Xu, and N. Zhang, "Trajectory planning algorithm for merging control of heterogeneous vehicular platoon on curve road," *Transport,* vol. 37, no. 4, pp. 279-290, 2022.

[28] M. S. Rahman, and M. Abdel-Aty, "Longitudinal safety evaluation of connected vehicles' platooning on expressways," *Accident Analysis and Prevention,* vol. 117, pp. 381-391, Aug, 2018.

[29] A. Shetty, H. Tavafoghi, A. Kurzhanskiy, K. Poolla, and P. Varaiya, "Risk assessment of autonomous vehicles across diverse driving contexts, " *in 2021 IEEE International Intelligent Transportation Systems Conference (ITSC)*, pp. 712-719.

[30] W. Hu, L. Y. Kang, and Z. G. Yu, "A possibilistic risk assessment framework for unmanned electric vehicles with predict of uncertainty traffic," *Frontiers in Energy Research,* vol. 10, Jun.2022.

[31] J. Y. Han, J. Zhao, B. Zhu, and D. J. Song, "Spatial-temporal risk field for intelligent connected vehicle in dynamic traffic and application in trajectory planning," *IEEE Transactions on Intelligent Transportation Systems,* vol. 24, no. 3, pp. 2963-2975, Mar.2023.

[32] M. Jiang, T. Wu, Z. Wang, Y. Gong, L. Zhang, and R. P. Liu, "A multi-intersection vehicular cooperative control based on end-edge-cloud computing," *IEEE Transactions on Vehicular Technology,* vol. 71, no. 3, pp. 2459-2471, 2022.

[33] H. Xu, Z. Cai, R. Li, and W. Li, "Efficient citycam-to-edge cooperative learning for vehicle counting in ITS," *IEEE Transactions on Intelligent Transportation Systems,* vol. 23, no. 9, pp. 16600-16611, 2022.

[34] X. Xu, M. Lai, H. Zhang, X. Dong, T. Li, J. Wu, Y. Li, and T. Peng, "Spatio-temporal cooperative control method of highway ramp merge based on vehicle-road coordination, " *in 2024 12th International Conference on Traffic and Logistic Engineering (ICTLE)*, pp. 93-98.

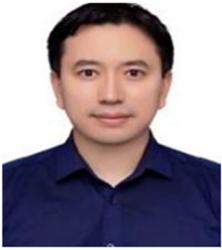

**Ting PENG** (S'06-M'08) Ph.D. in Computer Science from Xi'an Jiaotong University, is an Associate Professor at Highway School of Chang'an University. His research focuses on intelligent transportation systems, infrastructure monitoring, digital twins in highway engineering, highway maintenance, and civil engineering materials. He has led or participated in seven national research projects. He holds nine patents, over ten software copyrights, a National Science and Technology Progress Award, and a Provincial Science and Technology Award. He has published dozens of academic papers in journals and conferences.

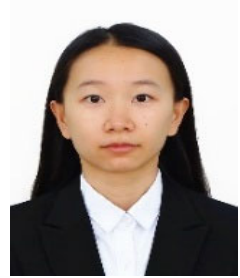

**Xiang DONG** is a postgraduate student at the Highway School of Chang'an University. She received her B.S. degree in Transportation Engineering from the City College of Southwest University of Science and Technology in 2022. Her current research interests include intelligent transportation systems and engineering data analysis.

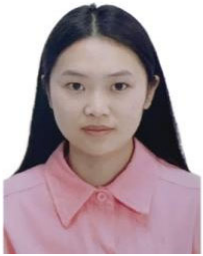

**Xiaoxue XU** is a postgraduate student at the Highway School of Chang'an University. She received her B.S. degree in Transportation Engineering from the City College of Southwest University of Science and Technology in 2023. Her current research interests include intelligent transportation systems and engineering data analysis.

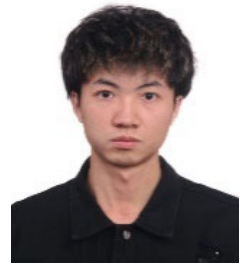

**Yincai CAI** is a postgraduate student at the Highway School of Chang'an University. His current research interests include intelligent detection technology for infrastructure and engineering data analysis.

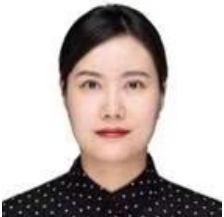

**Yuan LI** Ph.D. in Mechanics from Xi'an Jiaotong University, is a lecturer at Chang'an University. Her research focuses on intelligent transportation systems, multi-scale destruction of solid materials and multi-field coupled mechanics. She has led and participated in four nationally funded projects, including those from the National Natural Science Foundation of China, the Ministry of Education, and the Central University Foundation, and has published numerous academic papers in both domestic and international journals.

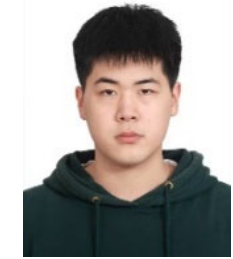

**Peng WU** is a postgraduate student at the Highway School of Chang'an University. He received his B.S. degree in Civil Engineering from Chongqing Jiaotong University in 2022. His current research interests include intelligent detection technology for infrastructure and engineering data analysis.

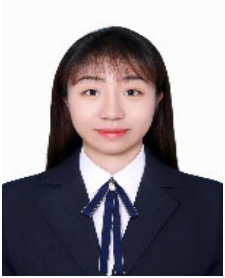

**Jie WU** is a postgraduate student at the School of Highway, Chang'an University. She received her Bachelor of Science degree in Traffic Engineering from Tongji Zhejiang College in 2021 and her Master of Science degree in Transportation from Chang'an University in 2024. Her research focuses on intelligent transportation systems.

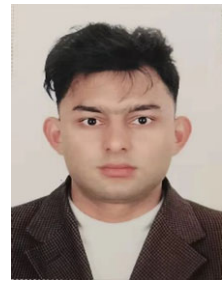

**Sana Ullah** is a postgraduate student at the Highway School of Chang'an University. He received his B.S. degree in Civil Engineering from Yunnan Technology & Business University in 2023. His current research interests include intelligent detection technology for infrastructure and engineering data analysis.

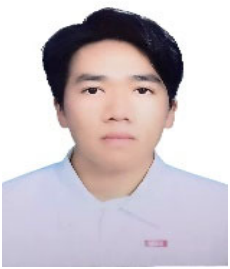

**Tao LI** is a postgraduate student at the Highway School of Chang'an University. He received his B.S. degree in Road Bridge and River-Crossing Engineering from Guangdong University of Technology in 2023. His current research interests include intelligent transportation systems, intelligent detection technology for infrastructure and engineering data analysis.